\documentclass[aps,reprint,twocolumns,pra,superscriptaddress,floatfix,notitlepage,nofootinbib]{revtex4-1}
\usepackage{amssymb,amsmath,amsfonts} 
\usepackage{mathtools}
\usepackage{physics}
\usepackage{array}
\usepackage{graphicx,epsfig,xcolor}
\usepackage[colorlinks=true,citecolor=blue,linkcolor=red]{hyperref}
\usepackage{newtxtext,newtxmath}

\newcommand*{\httpslink}[2][]{
  \ifthenelse{\equal{#1}{}}%
    {\href{https://#2}{#2}}%
    {\href{https://#2}{#1}}}

\newcommand{\beq}{\begin{equation}}
\newcommand{\eeq}{\end{equation}}

\begin{document}

\title{Comment on `Revisiting the phase transitions of the Dicke model'}

\author{\'{A}ngel L. Corps}
    \email[]{corps.angel.l@gmail.com}
    \affiliation{Instituto de Estructura de la Materia, IEM-CSIC, Serrano 123, E-28006 Madrid, Spain}
    \affiliation{Grupo Interdisciplinar de Sistemas Complejos (GISC),
Universidad Complutense de Madrid, Avenida Complutense s/n, E-28040 Madrid, Spain}
    
\author{Armando Rela\~{n}o}
    \email[]{armando.relano@fis.ucm.es}
    \affiliation{Grupo Interdisciplinar de Sistemas Complejos (GISC),
Universidad Complutense de Madrid, Avenida Complutense s/n, E-28040 Madrid, Spain}
    \affiliation{Departamento de Estructura de la Materia, F\'{i}sica T\'{e}rmica y Electr\'{o}nica, Universidad Complutense de Madrid, Avenida Complutense s/n, E-28040 Madrid, Spain}

\date{\today} 

\begin{abstract}
    In the work of Das and Sharma [\httpslink[\textcolor{blue}{Phys. Rev. A \textbf{105}, 033716 (2022)}]{journals.aps.org/pra/abstract/10.1103/PhysRevA.105.033716}] the phase transitions of the Dicke model are studied. Its main result is that, besides the well-known quantum phase transition, excited-state quantum phase transition and thermal phase transition exhibited by the model, there exists an upper bound energy, $E_{*}$, beyond which the model ceases to exhibit quantum chaotic behavior and the structure of the eigenfunctions changes. Based on this finding, a number of well-established results about the Dicke model are called into question. We argue that this result and all its consequences are spurious numerical effects resulting from an improper truncation of the infinite-dimensional Hilbert space necessary for numerical diagonalization.
\end{abstract}

\maketitle

In Ref. \cite{Das2022} the various phase transitions of the Dicke model of quantum optics \cite{Dicke1954} are revisited. This model consists of a set of $N$ two-level atoms with constant level splitting $\hbar \omega_{0}$ and a monochromatic bosonic (electromagnetic) field of frequency $\omega$, interacting with an intensity given by a coupling constant $g$. Its Hilbert space is the direct product of the atomic and photonic spaces, $H=H_{\textrm{at}}\otimes H_{\textrm{ph}}$. The atomic space is finite-dimensional for all number of atoms, $N$, whereas the photonic Hilbert space is that of the harmonic oscillator and, thus, it is infinite-dimensional. Hence, the single extensive parameter of the model is the number of atoms, $N$, and computing any physical quantity requires considering the unbounded photonic space. For theoretical calculations, this can be explicitly done; as well-known examples, we highlight:

(1) The calculation of partition functions determining the thermal phase transition \cite{Hepp1973,Wang1973,Carmichael1973,Comer1974,Aparicio2012,Brandes2013,Bastarrachea2016,PerezFernandez2017}. In these works, the partition function is computed  by performing a finite sum in the atomic space and an \textit{infinite} sum in the photonic space, for any number of atoms, $N$. The result evidences a thermal phase transition in the \textit{thermodynamic limit}, i.e.,  $N\to\infty$. 

(2) The semiclassical calculations, see e.g. \cite{Brandes2013,Bastarrachea2014,Altland2012,Emary2003,Emary2003PRE}. In these semiclassical analyses, the atomic phase space is bounded (for example, the atomic classical variables  may be restricted to a 2-dimensional ball of radius 2). However, the photonic classical variables can take  any real value and, thus, they are unbounded. The resulting classical phase space is therefore also unbounded (a possible choice is $\mathcal{M}=\mathbb{S}^{2}\times \mathbb{R}^{2}\subset\mathbb{R}^{4}$). 

Yet, numerical calculations can only be achieved by \textit{truncating} the infinite-dimensional photonic space, effectively reducing its dimension down to a \textit{finite} value $n_{\max}+1$, where $n_{\max}\in\mathbb{N}$ is the number of photons incorporated in the photonic basis $\{\ket{n}\}$, $n=0,1,2,\ldots,n_{\max}<\infty$ \cite{Emary2003PRE}. Notwithstanding, one has to be very careful with this truncation. This point is clearly explained in \cite{PerezFernandez2011b}: '[the basis of the Dicke model] is therefore infinite and must be numerically truncated, \textit{making the convergence test of the diagonalization an important issue} [...] In practice, \textit{one has to perform several
runs of the computation with increasing value of $n_{\max}$ and
compare the results obtained in each run}' (notation has been adapted). 

The importance of this remark is clearly illustrated by the following scenario. Let us imagine that we perform numerical calculations to predict experimental results. For concreteness, let us have in mind the famous experimental realization of the Dicke model with a condensate of $N$ $^{87}$Rb atoms trapped in an optical cavity \cite{Baumann2010}, and let us suppose that we have no technological limitations and, hence, that we can control with arbitrary precision the number of atoms of the condensate, $N$, and also the atom-field coupling strength, $g$, through the laser frequencies and intensities \cite{Dimer2007}. The question we want to answer is: `What would we observe if we were to measure any physical observable, $\mathcal{O}$, in this experiment?'. Obviously, we have no experimental control of the maximum number of photons introduced in the previous paragraph, $n_{\textrm{max}}$, because this is \textit{not} a physical parameter of the system, but just a numerical trick needed to obtain results on a computer. Hence, because the photonic Hilbert space of the Dicke model is infinite-dimensional, the physics of the experiment is given by the limit $n_{\textrm{max}} \rightarrow \infty$, for any values of $N$ and $g$. Therefore, any numerical prediction must be calculated by extrapolating the result obtained with a finite value of $n_{\textrm{max}}$, ${\mathcal O}(n_{\textrm{max}})$, to the limit $n_{\textrm{max}} \rightarrow \infty$: ${\mathcal O}_{\textrm{observed}} = \lim_{n_{\textrm{max}}\rightarrow \infty} {\mathcal O}(n_{\textrm{max}})$.

\begin{center}
\begin{figure}[h!]
\hspace*{-0.8cm}\includegraphics[width=0.55\textwidth]{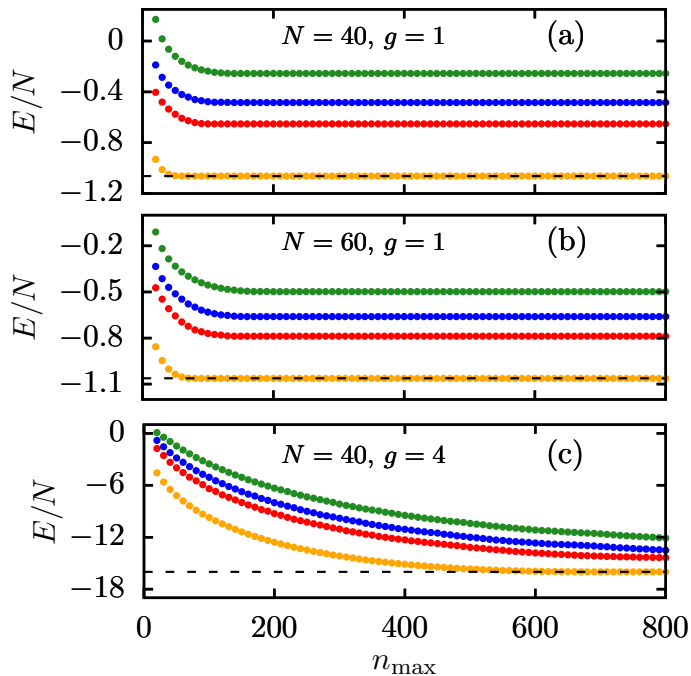}
\caption{Convergence of several positive-parity eigenlevels $E_{n}$ as a function of $n_{\max}$. In all panels, from bottom to top, the level index is $n=1,50,100,200$ ($n=1$ stands for the ground-state). (a-c) shows different choices of $N$ and $g$. Dashed horizontal lines represent the semiclassical ground-state energy \cite{Brandes2013,Bastarrachea2014}.}
\label{fig:convergence}
\end{figure}
\end{center}

In Fig. \ref{fig:convergence} we rely on numerics to illustrate this fundamental issue. We set $\omega=\omega_{0}=1$ throughout. We show here how a number of energy levels change as a function of the value $n_{\textrm{max}}$ used in the numerical calculations, for different values of $N$ and $g$. The main message from all three panels is that we can extrapolate their values in the limit $n_{\textrm{max}} \rightarrow \infty$ if we do the numerics with a value of $n_{\textrm{max}}$ large enough. It is also worth noting that, the larger the values of $N$, $g$ and the excitation energy, the larger the value of $n_{\textrm{max}}$ to predict what we would observe in an experiment. This remark is indeed well known \cite{PerezFernandez2011b}. Results in Fig. \ref{fig:convergence}(c) are especially relevant. Here, we work with $g=4$ and $N=40$. For $n_{\max}\sim 200$, the value used in Figs. 5(b) and 5(c) of \cite{Das2022} (with an even larger system size, $N=60$) \textit{no eigenlevel is properly converged}, not even the ground-state. It follows that the results of Figs. 2, where a huge system size, $N=512$, with only $n_{\textrm{max}}=32$ is used, 3, 4, 6 and 7 in \cite{Das2022} are incorrect: none of them would coincide with experimental results obtained with the same values of $N$ and $g$.

From this discussion, we spotlight the following conclusion: 

\textit{Any result depending on the maximum number of photons considered in the truncation of the photonic space, $n_{\max}$, is spurious, does not account for the real physics of the Dicke model, and hence cannot be observed in an experiment.}

The main result in \cite{Das2022}, the upper bound for the energy $E_{*}$, is an example of such a spurious effect. In particular, Eq. (16) in \cite{Das2022}, which is used to interpret Figs. 5, 6, 7, and 8, reads
\beq\label{Estar}
\frac{E_{*}}{N}=\frac{\omega}{N}n_{\max}+\frac{\omega_{0}}{2},
\eeq
and therefore is just a consequence of the truncation of the photonic space. In the actual Dicke model $E_{*}\propto n_{\max}\to\infty$. This means that we would see none of the main physical results discussed in \cite{Das2022} if we performed an experiment at an arbitrarily high excitation energy: neither the decrease of the degree of chaos, nor the decrease of the density of states, nor the decrease of the entanglement between the atoms and the bosonic field. 
Additionally, the truncated Hilbert space dimension, $N_{D}\equiv (N+1)(n_{\max}+1)$, is not a physical parameter of the model, and therefore it cannot be used to infer conclusions about the real Dicke model; yet, this is done in Fig. 2 of \cite{Das2022}. As a matter of fact, the effective Hilbert space dimension needed to get the eigenlevels up to a certain threshold converged depends on the basis used to build the Hamiltonian itself \cite{Bastarrachea2014basis,Hirsch2014}. 

Below we discuss some other erroneous conclusions obtained in \cite{Das2022}, as a consequence of the truncation of the photonic phase space.

(a) On the density of states (DOS). The Dicke model has a well-known excited-state quantum phase transition (ESQPT) \cite{Cejnar2021} at $E_{c1}/N=-1/2$. 
Additionally, the already existing literature reports the presence of a second critical energy (whether this may be considered as a true ESQPT is controversial), located exactly at $E_{c2}/N=1/2$ \cite{Brandes2013,Bastarrachea2014}. This is characterized by a non-analytic value of the energy, beyond which the DOS becomes constant, resulting from the fact that the finite-dimensional atomic Hilbert space becomes fully covered at this point. However, in \cite{Das2022} it is claimed that `this feature could be an artifact of the semiclassical analysis of earlier studies \cite{PerezFernandez2017,Brandes2013,Bastarrachea2014}'. This is a wrong conclusion. Owing to the \textit{collective} nature of the Dicke model, all semiclassical approaches involve setting up an effective Planck's constant $\hbar_{\textrm{eff}}\propto 1/N$. Therefore, the thermodynamic limit, $N\to\infty$, happens to coincide exactly with a semiclassical limit, $\hbar_{\textrm{eff}}\to 0$ (for a detailed discussion, see \cite{Cejnar2021}). Hence, the discrepancy between the semiclassical calculation and the numerical results reported in \cite{Das2022} is really due, again, to the improper truncation of the bosonic Hilbert space.

\begin{center}
\begin{figure}[h!]
\hspace*{0cm}\includegraphics[width=0.47\textwidth]{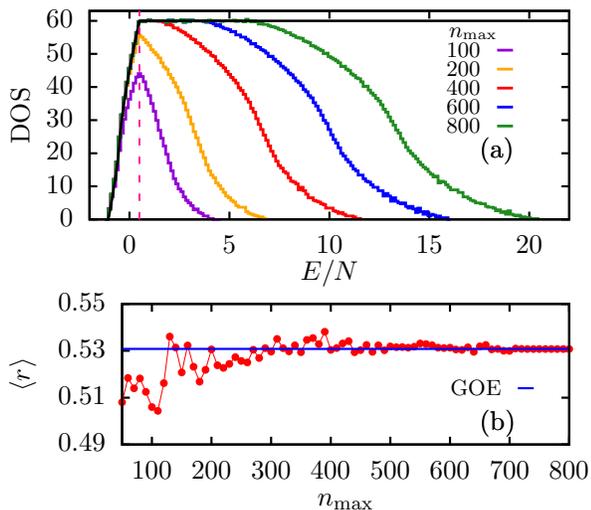}
\caption{(a) Density of states for the Dicke model with $g=1$, $N=60$ and several $n_{\max}$ (histograms). The black line represents the semiclassical result from \cite{Brandes2013,Bastarrachea2014}. The dashed horizontal line marks the energy $E/N=0.5$. (b) Average level gap ratio in the range $E/N\in[0.5,5.0]$ (points). The horizontal line represents the GOE theoretical result \cite{Atas2013,Corps2020}. }
\label{fig:densityrmean}
\end{figure}
\end{center}

We illustrate this statement in Fig. \ref{fig:densityrmean}(a), where we represent the DOS for $N=60$, $g=1$ and several values of $n_{\max}$. It is clearly observed that, as $n_{\max}$ increases, the DOS approaches the semiclassical limit result \cite{Brandes2013,Bastarrachea2014}, depicted with a black line, including within the flat region semiclassically predicted for $E/N>1/2$ \cite{Brandes2013,Bastarrachea2014}. Similar results were published in \cite{Bastarrachea2014}, where the convergence was stringently checked. It is worth noting that our result with $n_{\textrm{max}}=100$ is very similar to the one depicted in Fig. 8(b) of \cite{Das2022} with $g=2$; both of them show a maximum at $E/N=1/2$, with a decreasing density of states for $E/N>1/2$. Our results clearly show that this is again a spurious convergence issue. Furthermore, in \cite{Das2022} it is stated that `in the semiclassical approach of Brandes \cite{Brandes2013}, the upper cutoff is argued to be exactly at $\omega_0/2$, which would be consistent with our numerical data provided the limit $N \to \infty$ is taken before the $n_{\textrm{max}} \to \infty$' \cite{Das2022}. This sentence contains two important mistakes. First, it does not account for a non-analytical point in the density of states; rather, it concerns a spurious dynamical change due to convergence issues. Second, and more importantly, it considers $n_{\textrm{max}}$ as a physical parameter of the model, playing a role similar to $N$ when taking the thermodynamic limit. As pointed out at the beginning of this Comment, the number of photons $n_{\max}$ in the Dicke model is unbounded for any $N$. Therefore, an experimental measurement of the DOS would show a non-analytical point at $E/N=\omega_0/2$, and a flat behavior for higher energies, as demonstrated in \cite{Bastarrachea2014} and as shown in Fig. \ref{fig:densityrmean}(a) of this Comment.

As a complement, we represent in Fig. \ref{fig:densityrmean}(b) the average value of the ratio of consecutive level spacings as obtained from all eigenlevels in the range $E/N\in[0.5,5.0]$ for $g=1$, $N=60$ and different $n_{\max}$. We choose this region as \textit{all} eigenlevels within it are perfectly converged for $g=1,\,N=60,\,n_{\max}=800$.  
For a chaotic system, $\langle r\rangle$ is given by the result from the Gaussian orthogonal ensemble (GOE), $\langle r \rangle_{\textrm{GOE}}\approx 0.5307$ \cite{Atas2013,Corps2020}. It is clearly observed that $\langle r\rangle $ approaches $\langle r\rangle_{\textrm{GOE}}$ as $n_{\max}$ increases, and plateaus around this value at a given $n_{\max}$. This happens once the the spectrum is completely converged in the energy range considered, i.e., further increasing $n_{\max}$ does not alter the converged eigenvalues, and therefore the $r$ statistics remains unchanged. However, for small $n_{\max}$ not all eigenlevels in the considered range are converged, and therefore fluctuations are apparent in Fig. \ref{fig:densityrmean}(b) for small $n_{\max}$. Together with our previous result about the DOS, this implies that we can correctly predict the experimental results for both the DOS and $\langle r \rangle$ for $N=60$, $g=1$, and $E/N \leq 5$ by truncating the bosonic Hilbert space with $n_{\textrm{max}}=800$, but cannot reach higher energies in the same way because our numerical calculations would require an ever-growing $n_{\max}$.

(b) On quantum chaos. The Dicke model is known to be chaotic for $g>g_{c}\equiv \sqrt{\omega\omega_{0}}/2$ and high excitation energies  \cite{Chavez2016}; the low excitation spectrum for $g>g_{c}$ is approximately integrable due to the existence of an additional approximate integral of motion \cite{Relano2016,Bastarrachea2017}. Nevertheless, in \cite{Das2022} both of these well-established results are called into question. 

For the second one, a multifractal scaling of the ground-state wavefunctions is performed in terms of the spurious Hilbert space dimension of the arbitrarily truncated infinite-dimensional Hilbert space\footnote{It is possible to define an effective dimension of the Hilbert space in terms of the number of bosons required for convergence, as done in \cite{Hwang2015,Puebla2016} to define a proper thermodynamic limit for the Rabi model (a version of the Dicke model with just one two-level atom). In this way, the part of the infinite Hilbert space effectively explored by the system is taken as a real physical parameter. Yet, the method in \cite{Das2022} consists in arbitrarily fixing $n_{\textrm{max}}$, and changing only the number of atoms $N$. Therefore, not only is the scaling  performed in terms of the dimension of an arbitrarily truncated Hilbert space, but the eigenfunctions used in the calculations become less converged as the system size increases.}, $N_D$, and applied to improperly converged wavefunctions; these results are in Fig. 2 of \cite{Das2022}. 

\begin{center}
\begin{figure}[h!]
\hspace*{0cm}\includegraphics[width=0.49\textwidth]{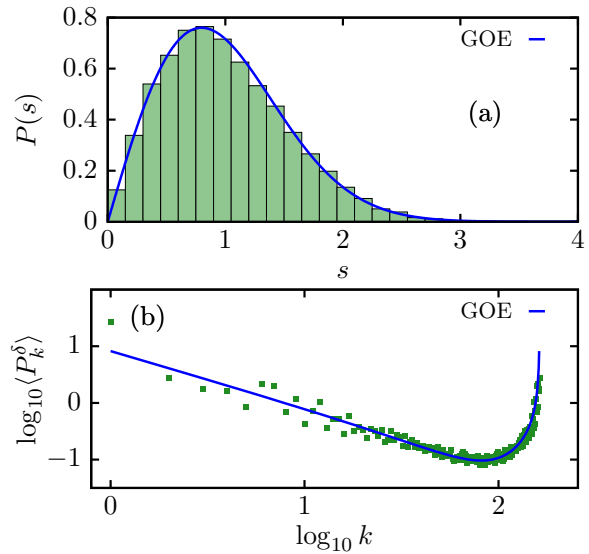}
\caption{(a) Level spacing distribution $P(s)$ in the high energy range $E/N\in[0.5,5.0]$ for $g=1$, $N=60$ and $n_{\max}=800$ (histogram). The blue line depicts the Wigner-Dyson distribution. (b) Power spectrum of the level motion \cite{Relano2002} for the same energy range (points). Results correspond to the average power spectrum as obtained from splitting the energy interval into 50 equal parts. The blue line represents the GOE theoretical result \cite{Faleiro2004}. }
\label{fig:spectral}
\end{figure}
\end{center}

For the first one, they propose an upper bound for the energy of the Dicke model, $E_*$, given in Eq. (16) in \cite{Das2022} and in Eq. \eqref{Estar} in this Comment; above $E_{*}$, the dynamics of the Dicke model is supposedly no longer chaotic. Furthermore, in \cite{Das2022} it is stated that `a horizontal portion [of the DOS] in the super-radiant phase would be inconsistent with Wigner-Dyson statistics'. As we have pointed out above, this vanishing of chaos would not be observed in any experiment. In Fig. \ref{fig:spectral}, we focus on the properly converged eigenlevels with $N=60$, $g=1$, and $E/N \in \left[0.5,5.0\right]$, for which the density of states is perfectly flat, as shown in Fig. \ref{fig:densityrmean}(a), to predict what we would really observe in such an experiment. We can see that these excited states of the Dicke model are indeed chaotic, as originally reported in \cite{PerezFernandez2011}. In Fig. \ref{fig:spectral}(a) we have represented the (nearest-neighbor) level spacing distribution, $P(s)$, where the unfolded spacings (see below for details about the unfolding procedure) $\{s_{n}\}_{n}$ are defined from an ordered set of eigenlevels $\{E_{1}\leq E_{2}\leq\ldots\}$ as $s_{n}=S_{n}/\langle S\rangle$, where $S_{n}=E_{n+1}-E_{n}\geq0$ is the consecutive level gap. $P(s)$ is shown to agree excellently with GOE statistics, i.e., with the Wigner-Dyson distribution $P(s)=\frac{\pi}{2}s e^{-\pi s^{2}/4}$. Moreover, in Fig. \ref{fig:spectral}(b) we represent the averaged power spectrum of the level motion \cite{Relano2002}, $\langle P_{k}^{\delta}\rangle$, which is an indicator of long-range level correlations. Our numerical result is very close to the GOE prediction \cite{Faleiro2004} for almost all frequencies $k$, indicating that the spectrum exhibits GOE correlations also for quite distant unfolded eigenlevels. 

Hence, our results (in concert with original results \cite{PerezFernandez2011}) dismiss the claim in \cite{Das2022} about the incompatibility of Wigner-Dyson spectral statistics with a flat density of states. Indeed, we find this claim to be a serious misinterpretation of the well-established theory of quantum chaos \cite{Haake2010,Borgonovi2016,Gomez2011}. In their seminal work of 1984, Bohigas, Giannoni and Schmit \cite{Bohigas1984} formulated the famous conjecture that  `spectra of time-reversal-invariant systems whose classical analogues are K systems show the same \textit{fluctuation} properties as predicted by GOE'. One of the key elements of the theory of quantum chaos is indeed the fluctuations of eigenlevels. To analyze spectral statistics, one first needs to `unfold' the spectrum, i.e.,  transform the eigenlevels $\{E_{n}\}_{n}$ into a dimensionless sequence, $\{\epsilon_n = \overline{\mathcal{N}}(E_n)\}_{n}$ \cite{Gomez2002,Corps2021}. The main effect of this transformation is to remove the contribution of the smooth part of the cumulative level density, $\overline{\mathcal{N}}(E_n)$, to the spectral statistics; this corresponds precisely to the semiclassical density of states \cite{Gutzwiller1967,Berry1977} shown to be flat for the Dicke model at $E/N>1/2$ \cite{Brandes2013,Bastarrachea2014}. Therefore, despite the interpretation in \cite{Das2022}, the shape of the semiclassical density of states plays absolutely no role in spectral statistics. Quite contrarily, this regular part has to be removed to properly obtain the fluctuations of energy levels, which  constitute the main signature of quantum chaos and are at the core of this theory \cite{Bohigas1984}. 

Furthermore, the claim `the literature has many studies \cite{PerezFernandez2011,Lewis2019} of level statistics that separately look at energy levels below $E_{c}$ and above $E_{c}$, but with no mention of the upper cutoff. However, as our data show, the inclusion of the higher band energies in their level statistics study would make their Wigner-Dyson results noisy' \cite{Das2022} also is unsubstantiated. Results for Fig. 5(b,inset) in \cite{Das2022} are obtained with $g=4$, $N=60$, $n_{\max}=200$, and $E/N>3.8$. As clearly seen in Fig. \ref{fig:convergence}(c) of this Comment, not even the ground state energy is well converged with $n_{\max}=200$, $g=4$, and even a smaller value of $N=40$. Hence, this spurious deviation from chaos would not be observed in any experiment, irrespective of the energy at which the measurement were performed. 

In summary, in this Comment we have argued that the results reported in \cite{Das2022} suffer from important problems resulting from an incorrect truncation of the infinite-dimensional Hilbert space of the Dicke model as well as misconceptions about the meaning of a semiclassical approach and the theory of quantum chaos. Therefore, many of the conclusions there presented are essentially flawed and do not account for the real physics of the Dicke model. 

\begin{acknowledgments}
This work has been supported by the Spanish grant PGC-2018-094180-B-I00 funded by Ministerio de Ciencia e Innovaci\'{o}n/Agencia Estatal de Investigaci\'{o}n MCIN/AEI/10.13039/501100011033 and FEDER "A way of making Europe". A. L. C. acknowledges financial support from `la Caixa' Foundation (ID 100010434) through the fellowship LCF/BQ/DR21/11880024.
\end{acknowledgments}


\begin{thebibliography}{100}

\bibitem{Das2022} P. Das and A. Sharma, \textit{Revisiting the phase transitions of the Dicke model}, Phys. Rev. A \textbf{105}, 033716 (2022).

\bibitem{Dicke1954} R. H. Dicke, \textit{Coherence in Spontaneous Radiation Processes}, 
Phys. Rev. \textbf{93}, 99 (1954).

\bibitem{Hepp1973} K. Hepp and E. H. Lieb, \textit{On the superradiant phase transition for molecules in a quantized radiation field: the dicke maser model}, Ann. Phys. (NY) \textbf{76}, 360 (1973).

\bibitem{Wang1973} Y. K. Wang and F. T. Hieo, \textit{Phase Transition in the Dicke Model of Superradiance},  Phys.Rev. A \textbf{7}, 831 (1973).

\bibitem{Carmichael1973} H. J. Carmichael, C. W. Gardiner, and D. F. Walls, \textit{Higher order corrections to the Dicke superradiant phase transition}, Phys. Lett. A {\bf 46}, 47 (1973).

\bibitem{Comer1974} G. Comer Duncan, \textit{Effect of antiresonant atom-field interactions on phase transitions in the Dicke model}, Phys. Rev. A {\bf 9}, 418 (1974).

\bibitem{Aparicio2012} M. Aparicio Alcalde, M. Bucher, C. Emary, and T. Brandes, \textit{Thermal phase transitions for Dicke-type models in the ultrastrong-coupling limit}, Phys. Rev. E {\bf 86}, 012101 (2012).

\bibitem{Brandes2013} T. Brandes, \textit{Excited-state quantum phase transitions in Dicke superradiance models}, Phys. Rev. E \textbf{88}, 032133 (2013).

\bibitem{Bastarrachea2016} M. A. Bastarrachea-Magnani, S. Lerma-Hern\'andez, and J. G. Hirsch, \textit{Thermal and quantum phase transitions in atom-field systems: a microcanonical analysis}, J. Stat. Mech. (2016) 093105.

\bibitem{PerezFernandez2017} P. P\'{e}rez-Fern\'{a}ndez and A. Rela\~{n}o, \textit{From thermal to excited-state quantum phase transition: The Dicke model}, Phys. Rev. E \textbf{96}, 012121 (2017).

\bibitem{Bastarrachea2014} M. A. Bastarrachea-Magnani, S. Lerma-Hern\'{a}ndez, and J. G. Hirsch, \textit{Comparative quantum and semiclassical analysis of atom-field systems. I. Density of states and excited-state quantum phase transitions}, Phys.Rev. A \textbf{89}, 032101 (2014).

\bibitem{Altland2012} A. Altland and F. Haake, \textit{Equilibration and macroscopic quantum fluctuations in the Dicke model}, New J. Phys. \textbf{14} 073011 (2012).

\bibitem{Emary2003} C. Emary and T. Brandes, \textit{Quantum Chaos Triggered by Precursors of a Quantum Phase Transition: The Dicke Model}, 
Phys. Rev. Lett. \textbf{90}, 044101 (2003).

\bibitem{Emary2003PRE} C. Emary and T. Brandes, \textit{Chaos and the quantum phase transition in the Dicke model}, Phys. Rev. E \textbf{67}, 066203 (2003).

\bibitem{PerezFernandez2011b} P. P\'{e}rez-Fern\'{a}ndez, P. Cejnar, J. M. Arias, J. Dukelsky, J. E. Garc\'{i}a-Ramos, and A. Rela\~{n}o, \textit{Quantum quench influenced by an excited-state quantum phase transition}, Phys. Rev. A \textbf{83}, 033802 (2011).

\bibitem{Baumann2010} K. Baumann, C. Guerlin, F. Brennecke, and T. Esslinger, \textit{Dicke quantum phase transition with a superfluid gas in an optical cavity}, Nature {\bf 464}, 1301 (2010).

\bibitem{Dimer2007} F. Dimer, B. Estienne, A. S. Parkins, and H. J. Carmichael, \textit{Proposed realization of the Dicke-model quantum phase transition in an optical cavity QED system}, Phys. Rev. A {\bf 75}, 013804 (2007).

\bibitem{Bastarrachea2014basis} M. A. Bastarrachea-Magnani and J. G. Hirsch, \textit{ Efficient basis for the Dicke model: I. Theory and convergence in energy,} Phys. Scr. \textbf{2014} 014005.

\bibitem{Hirsch2014} J. G. Hirsch and M. A. Bastarrachea-Magnani, \textit{Efficient basis for the Dicke model: II. Wave function convergence and excited states}, Phys. Scr. \textbf{2014} 014018.

\bibitem{Cejnar2021} P. Cejnar, P. Str\'{a}nsk\'{y}, M. Macek, and M. Kloc, \textit{Excited-state quantum phase transitions}, J. Phys. A: Math. Theor. \textbf{54}, 133001 (2021).

\bibitem{Atas2013} Y. Y. Atas, E. Bogomolny, O. Giraud and G. Roux, \textit{Distribution of the Ratio of Consecutive Level Spacings in Random Matrix Ensembles}, Phys. Rev. Lett. \textbf{110}, 084101 (2013).

\bibitem{Corps2020} A. L. Corps and A. Rela\~{n}o, \textit{Distribution of the ratio of consecutive level spacings for different symmetries and degrees of chaos}, Phys. Rev. E \textbf{101}, 022222 (2020).

\bibitem{Chavez2016} J. Ch\'avez-Carlos, M. A. Bastarrachea-Magnani, S. Lerma-Hern\'andez, and J. G. Hirsch, \textit{Classical chaos in atom-field systems}, Phys. Rev. E {\bf 94}, 022209 (2016).

\bibitem{Relano2016} A. Rela\~no, M. A. Bastarrachea-Magnani, and S. Lerma-Hern\'andez, \textit{Approximated integrability of the Dicke model}, EPL {\bf 116} (2016), 50005.

\bibitem{Bastarrachea2017} M. A. Bastarrachea-Magnani, A. Rela\~{n}o, S. Lerma-Hern\'andez, B. L\'opez-del-Carpio, J. Ch\'avez-Carlos, J. G. Hirsch, \textit{Adiabatic invariants for the regular region of the
Dicke model}, J. Phys. A: Math. Theor. {\bf 50} (2017) 144002.

\bibitem{Hwang2015} M.-J. Hwang, R. Puebla, and M. B. Plenio, \textit{Quantum Phase Transition and Universal Dynamics in the Rabi Model}, Phys. Rev. Lett. {\bf 115}, 180404 (2015).

\bibitem{Puebla2016} R. Puebla, M.-J. Hwang, and M. B. Plenio, \textit{Excited-state quantum phase transition in the Rabi model}, Phys. Rev. A {\bf 94}, 023835 (2016).

\bibitem{PerezFernandez2011} P. P\'{e}rez-Fern\'{a}ndez, A. Rela\~{n}o, J. M. Arias, P. Cejnar, J. Dukelsky, and J. E. Garc\'{i}a-Ramos, \textit{Excited-state phase transition and onset of chaos in quantum optical models,}
Phys. Rev. E \textbf{83}, 046208 (2011).

\bibitem{Relano2002} A. Rela\~{n}o, J. M. G. G\'{o}mez, R. A. Molina, and E. Faleiro, \textit{Quantum chaos and 1/f noise}, Phys. Rev. Lett. \textbf{89}, 244102 (2002).

\bibitem{Faleiro2004} E. Faleiro, J. M. G. G\'{o}mez, R. A. Molina. L. Mu\~{n}oz, A. Rela\~{n}o, and J. Retamosa, \textit{Theoretical derivation of 1/f noise in quantum chaos}, Phys. Rev. Lett. \textbf{93}, 244101 (2004).

\bibitem{Haake2010} F. Haake, Quantum Signatures of Chaos (Berlin, Springer).

\bibitem{Borgonovi2016} F. Borgonovi, F. M. Izrailev, L. F. Santos and V. G. Zelevinsky, \textit{Quantum Chaos and Thermalization in Isolated Systems of Interacting Particles}, Phys. Rep. \textbf{626}, 1-58 (2016).

\bibitem{Gomez2011} J. M. G. G\'{o}mez, K. Kar, V. K. B. Kota, R. A. Molina, A. Rela\~{n}o, and J. Retamosa, \textit{Many-body quantum chaos: Recent developments and applications to nuclei}, Phys. Rep. \textbf{499}, 103-226 (2011).

\bibitem{Bohigas1984} O. Bohigas, M. J. Giannoni and C. Schmit, \textit{Characterization of chaotic quantum spectra and universality of level fluctuation laws}, Phys. Rev. Lett. \textbf{52}, 1 (1984).

\bibitem{Gomez2002} J. M. G. G\'{o}mez, R. A. Molina, A. Rela\~{n}o, and J. Retamosa, \textit{Misleading signatures of quantum chaos}, Phys. Rev. E \textbf{66}, 036209 (2002).

\bibitem{Corps2021} A. L. Corps and A. Rela\~{n}o, \textit{Long-range level correlations in quantum systems with finite Hilbert space dimension}, Phys. Rev. E \textbf{103}, 012208 (2021).

\bibitem{Gutzwiller1967} M. C. Gutzwiller, \textit{Phase-integral approximation in momentum space and the bound states of an atom}, J. Math. Phys. {\bf 8}, 1979 (1967); M. C. Gutzwiller, \textit{Phase-integral approximation in momentum space and the bound states of an atom II}, J. Math. Phys. {\bf 10}, 1004 (1969); M. C. Gutzwiller, \textit{Energy spectrum according to classical mechanics}, J. Math. Phys. {\bf 11}, 1791 (1970); M. C. Gutzwiller, \textit{Periodic orbits and classical quantiztion conditions}, J. Math. Phys. {\bf 12}, 343 (1971).

\bibitem{Berry1977} M. V. Berry and M. Tabor, \textit{Calculating the bound spectrum by path summation in action-angle variables}, J. Phys. A {\bf 10}, 371 (1977).

\bibitem{Lewis2019} R. Lewis-Swan, A. Safavi-Naini, J. J. Bollinger, and A. M. Rey, \textit{Unifying scrambling, thermalization and entanglement through measurement of fidelity out-of-time-order correlators in the Dicke model}, Nat. Commun. \textbf{10}, 1581 (2019).


\end{thebibliography}
\end{document}